# Model versions and fast algorithms for network epidemiology

Petter Holme

Department of Energy Science, Sungkyunkwan University, Suwon 440-746, Korea
IceLab, Department of Physics, Umeå University, 90187 Umeå, Sweden
Department of Sociology, Stockholm University, 10691 Stockholm, Sweden

**Abstract**

Network epidemiology has become a core framework for investigating the role of human contact patterns in the spreading of infectious diseases. In network epidemiology represents the contact structure as a network of nodes (individuals) connected by links (sometimes as a temporal network where the links are not continuously active) and the disease as a compartmental model (where individuals are assigned states with respect to the disease and follow certain transition rules between the states). In this paper, we discuss fast algorithms for such simulations and also compare two commonly used versions—one where there is a constant recovery rate (the number of individuals that stop being infectious per time is proportional to the number of such people), the other where the duration of the disease is constant. We find that, for most practical purposes, these versions are qualitatively the same.

**Introduction**

To understand the spreading of infectious diseases in populations, one typically uses compartmental models [1]. Such models divide people into classes with respect to a disease—for example susceptible (S), infectious (I), or recovered (R)—and assign transition rules between these classes. The two canonical models of this type are the Susceptible–Infectious–Recovered (SIR) and Susceptible–Infectious–Susceptible (SIS) models. In the SIR model a susceptible (S) individual can, upon meeting an infectious individual become infectious. After a while the infectious recovers—i.e. goes from I to R. In the SIS model, the infectious becomes susceptible again, rather than recovered. Symbolically these processes can be described as:

$$\text{SIR model:} \begin{cases} S, I \to I, I \\ I \to R \end{cases} \qquad (1)$$

and

$$\text{SIS model:} \begin{cases} S, I \to I, I \\ I \to S \end{cases} \qquad (2)$$

For completeness, one also needs to specify when and between whom contacts happen and the conditions for a transition between states to happen. In the general case, one needs to represent contacts as a temporal network [2,3], e.g. a *contact sequence* listing which pair of individuals ($i,j$) that have been in contact at a (discrete) time step $t$. One often assumes system not to have any structure in the time dimension, so that the system is well modeled by a static network and a constant contact rate across the links [4–6]. In most cases—whether a static or temporal network—one cannot do much better than assuming that the disease spreads to a susceptible with a fixed probability λ per SI contact. For the cessation of the I state one does, however, have some choices. The traditional approach is to assume that all infectious individuals become removed (or re-susceptible) with a constant probability per unit time. In an infinite network, this means that the time in the I state is exponentially distributed, which disagrees with medical data [7]. However, it could be an acceptable approximation since it simplifies the model much, especially in differential equation modeling. An alternative approach is to let the duration δ of the infectious stage, rather than the recovery rate, to be constant. The goal of this paper is to present and compare these two approaches to the SIR model on disease spreading on static networks. Other compartmental models could be handled similarly. We will mention some special considerations for the SIS model. We will discuss the algorithms, their difference in output and what this tells us about the modeling of disease spreading.

**SIR and SIS models for simulations on networks**

In this section we discuss the detailed algorithms for compartmental models on networks. C code for the three algorithms for the constant duration version can be found at https://github.com/pholme/sir_algos.

*Constant disease duration, canonical algorithm*

In the version of the compartmental models where the duration of the infectious stage is held constant, one need a first-in-first-out data structure to efficiently keep track of the infectious individuals. This is essentially a one-dimensional list of values ordered by the times of infection. A practical such data structure (that is readily available for e.g. Python) is the double-ended queue, or "deque" for short. One basic operation to a deque is to append a value (in our case a pair—a node and its time of infection) to one side ("left") or the other ("right"). The other basic operation is to pop a value from left or right. This is the opposite operation to append where the value is read to a variable and deleted from the deque.

In languages like C or C++, with explicit (and more cumbersome) memory allocation, one can exploit the fact that this data structure does not have more elements than the number of individuals, $N$. One would allocate an array of $N$ elements and use two pointers to mark the beginning and end of the queue. In the SIS model, one may need to append more than $N$ values (as an individual can be re-infected). Once the pointer reaches the end of the array, it would wrap around to the beginning.

In Fig. 1, we show Python-flavored pseudo code for this algorithm. The code represents the body of a routine that runs one outbreak of the SIR model. The main loop is over the discrete time steps $t$. For the SIR model, the loop can be infinite (or set to its theoretical maximum $N\delta$). For the SIS model, the disease could live forever, so one needs some explicit cutoff to the iterations.

The first block of code in the main loop is to change individuals who have been infectious for δ time steps to recover (or, in the SIS model, susceptible). With a deque, one would pop elements from the list (in a while loop) until one finds an individual that is not yet recovered. Then one put that individual back, and breaks the loop. The second block of code checks if the process above emptied the list of infectious. If so, the outbreak is dead and the routine can end. The third block of code scans for infection events. One cannot immediately denote the newly infectious as infectious since that would allow the disease to spread further than one link during a time step. Instead we assign those nodes a temporary state ("to be infected"), store them in an array, and finally change all nodes in the array to infectious.

*Constant recovery rate*

The main idea behind the constant recovery rate version is that the classic SIR (or Kermack–McKendrick) model is a special case (for a fully connected network in the large-$N$ limit) [1]:

$$\frac{dS}{dt} = -\beta SI \tag{3}$$

$$\frac{dI}{dt} = \beta SI - \nu I \tag{4}$$

$$\frac{dR}{dt} = \nu I \tag{5}$$

Here $S$, $I$, $R$ are the number of susceptible, infectious and recovered, respectively; $\beta$ and $\nu$ are parameter values governing the infection and recovery rates. This classic formulation sets the timescale between contagion and recovery events, which we can exploit in our code. To be precise, we can set up the simulation so that, each iteration of the main loop, either a contagion or a recovery event occurs. For a simulation of a well-mixed population, we could obtain the number of infection events per time unit relative to the total number of infection and recovery events by combining Eq. (3) by Eq. (5), to achieve

$$\frac{\beta SI}{\nu I + \beta SI} = \frac{R_0 S}{1 + R_0 S} \tag{6}$$

where $R_0 = \beta/\nu$ is a classic parameter for the SIR model called the basic reproductive number. In the classical theory, starting with the Eqs. (3), (4) and (5), $R_0$ determines the final outbreak size $\Omega$ (more precisely, the fraction of recovered individuals when the infection has died). In particular, if $R_0 < 1$, then $\Omega = 0$, and for $R_0 > 1$ one have $\Omega > 0$. This threshold behavior also holds for the SIS model (but then $\Omega$ should be understood as the number of infection events per individual). $R_0$ has a more verbal definition (that coincides with the above for well-mixed populations)—it is the expected number of others an infectious individual would directly infect in an uninfected population. In this paper, we just use $R_0$ to abbreviate $\beta/\nu$. Eq. (6) tells us that we do not need to use $\nu$ and $\beta$ separately, but could use $R_0$ as our only model parameter. However, then we lose the possibility to study the dynamics in the true time—if we, for example, increase both $\nu$ and $\beta$ by a factor two, the disease would affect as many people, but run twice as fast through the population. Note that there is never a need for rejecting an infection or recovery event—one only need to use the algorithm outlined in this section, and recalculate the timescale if needed.

Simulating the SIR or SIS model on networks one has to modify Eq. (6) slightly. Note that $SI$ of the term $\beta SI$ (in Eqs. (3) and (4)) represents the number of possible infection events. In a not fully connected network, this is rather $M_{SI}$—the number of links between $S$ and $I$ individuals. So we need to change Eq. (6) to

$$\frac{R_0 M_{SI}}{I + R_0 M_{SI}} \tag{7}$$

So also in this case, it is enough for most purposes to consider $R_0$ as the only model parameter. The algorithm is outlined in Fig. 2.

*Further speed-up of the constant rate version via SI-link lists*

One can make the code outlined above yet a bit faster (but only a linear improvement) by storing the SI links in an array. Then, instead of going through the neighbors of infectious individuals to simulate potential infection events (which would be unnecessary if the neighbor is I or R), one would

```
# Initialize.
for all nodes i:
    mark i susceptible
mark source infectious

append source to the infectious_list
set source's infection time to 0

for time starting at 1:
    while the earliest infected node i in infectious_list…
                …has been infectious longer than δ:
        remove i from infectious_list
        mark i as recovered

    # Check if the outbreak died.
    if infectious_list is empty:
        return

    # Go through the SI contacts.
    empty to_change_to_infectious
    for nodes i in infectious_list:
        for neighbors j of i:
            if j is susceptible:
                with a probability λ:
                    append j to to_change_to_infectious
                    mark j as infectious next time

        # Change the ones to be infectious next time step.
        for nodes i in to_change_to_infectious:
            mark i as infectious
            append i to infectious_list
```

Fig. 1. Algorithm and Python flavored pseudo code for the SIR model with constant disease duration. Boldface indicates typical Python expressions. Text in italics represents descriptions that are not from any programming language. Many parts of the code are not shown, e.g. the libraries to include. λ, δ, the infection source, and the network itself are input to the routine; measurements and return values are not shown.

```
# Initialize.
for all vertices i:
    mark i susceptible
mark source infectious

initialize infectious_list to a list containing only source
initialize si_links to an empty list
for all neighbors i of source:
    append (i,source) to si_links

while there are SI links:
    # Calculate the probability of an infection event.
    prob_si = r0 * M_SI / (r0 * M_SI + I)

    with a probability prob_si: # SI → II code starts here.
        let (i,j) be a random element from si_links (with j
            being infectious)
        add i to infectious_list
        mark i infectious
        # Add new SI links, and delete new II links.
        for all neighbors j of i:
            if j is marked susceptible:
                append (i,j) to si_links
            elif j is marked infectious:
                delete (i,j) from si_links
    else: # I → R code starts here.
        # Recover a random infectious.
        let i be a random element from infectious
        delete i from infectious
        mark i recovered
        # Delete the new SR links from the list of SI links.
        for all neighbors j of i:
            if j is marked susceptible:
                delete (i,j) from si_links
```

Fig. 2. Pseudo code for the canonical SIR model with constant recovery rate. The notations are similar to Fig. 1. $R_0$ and the infection source are input parameters.

go through the SI links. To do this efficiently, one needs to delete links from the SI array without scanning it through. This can be done by a link structure that stores information about links, including their node id-numbers and the indices of the links in the SI-link array. Whenever an individual *i* recovers, one would delete the new IR or SR links by going through *i*'s neighbors, check their index in the SI-link list and delete it. This deletion involves replacing the link in question by the last element of the SI-link list and updating the link structure for that (previously last) element. A drawback is that using hash tables like this makes the code less readable.

*The no-state-label algorithm of the constant duration version*

Also for the constant duration case, one can speed up the calculation by avoiding checking whether the neighbor of an infectious node is infectious. Instead of an SI-link list, one can sort the nodes' neighbor lists so it starts with the susceptible neighbors, and keep a count of the number of such neighbors. To avoid having to search through the neighbor lists, each node needs to keep a list of its index in the corresponding neighbor's neighbor list (so say *i* and *j* are neighbors and *i* is neighbor number 3 in *j*'s neighbor list while *j* is neighbor number 2 in *i*'s neighbor list, then this index list of *i* would have the value 3 at position number 2). This procedure, along with the rest of the algorithm, is sketched in Fig. 3. Deleting a newly infectious node from a list of susceptible is then a constant time operation, and one also avoids labeling the nodes by their disease state. However, it is not always this algorithm beats the canonical one above. Depending on the network structure and parameter values, the bookkeeping (updating the list of indices, etc.) may outweigh the benefit from not having to check the state of all neighbors of infectious nodes.

*The incomplete SI-link-list algorithm of the constant duration version*

Our third algorithm for the constant duration SIR model takes an intermediate path between avoiding checking the state of neighbors of infectious nodes and avoiding extensive bookkeeping. Just like the fast algorithm for the constant rate version, it keeps a list of SI links, but allows II and SR links to appear in this list. Essentially, it keeps the IR links that are newly generated by recovery process until one runs the block of code representing infection events. It also keeps some II links (the ones where the susceptible got infected by someone else than the other end of the link during the last time step). It is convenient to use two of these SI lists alternately through the iterations, so during one time step one goes through the first list and fills up the second, the next time step one goes through the second list and fills up the first. This algorithm is outlined in Fig. 4.

*Running time comparison of the algorithms*

In Fig. 5 we compare the running times of the two alternative algorithms to the canonical one for various networks, sizes and parameter values. For all these simulations the average outbreak size is around $0.75N$. In panels A–C, we use random regular graphs with degree (number of neighbors) three. These are graphs where all nodes have the same degree, but the links are as random as possible. In panels D–F, we use Barabási–Albert networks [8] of average degree 20. The latter class of networks has a very different network structure—it is not only much denser on average, but has a broad distribution of degrees.

Our first observation is that there is much time to be saved with an appropriate choice of algorithm. This is especially true for the smaller system sizes where we observe running time less than half of the canonical algorithm. Furthermore, the incomplete link list algorithm is usually the fastest (for the random regular graphs it is consistently so). An optimal hybrid of these algorithm would not only chose the fastest of these three algorithms for the given graph structure and model parameters, but also shift procedure within a simulation run of a single outbreak. Presumably the canonical algorithm is comparatively fast in the beginning when there are not so many links ignored when the algorithm sweeps all the neighbors of infectious nodes. A fast hybrid algorithm could thus run the canonical

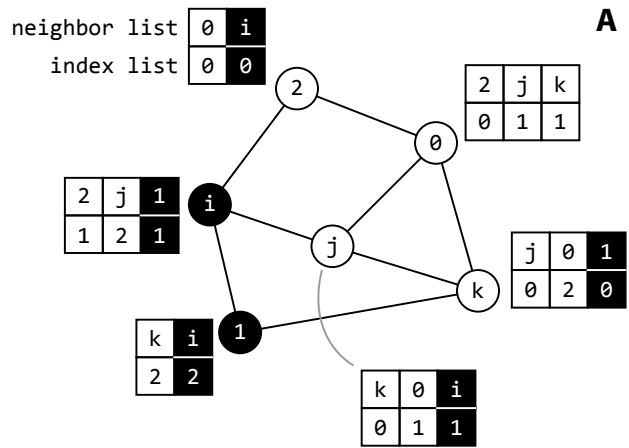

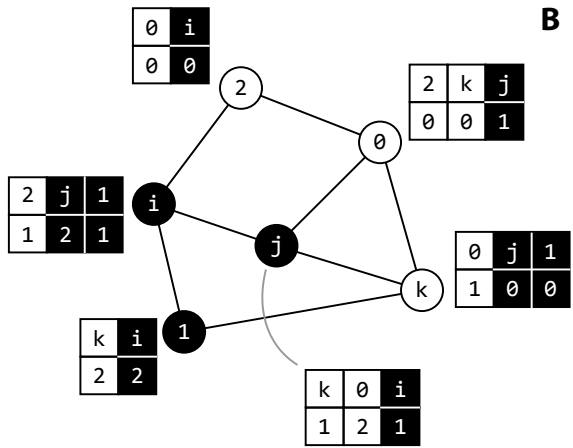

```
def index_swap(i):                                    C
    for all neighbors j of i:
        decrement j.ns
        j's index of k = j.ns
        k = j.nb[j's index of k]
        j's index of i = i.inx[i's index of j]
        k's index of j = j.inx[j's index of k]
        k.inx[k's index of j] = j's index of i
        i.inx[i's index of j] = j's index of k
        j.nb[j's index of k] = i
        j.nb[j's index of i] = k
        j.inx[j's index of k] = i's index of j
        j.inx[j's index of i] = k's index of j
```

```
                                                      D
# Initialize.
for all vertices i:
    label i susceptible
label source infectious

initialize infectious_list to a list containing only source
initialize source's infection time to 0
index_swap(source)

for time starting from 1:
    # I → R code starts here.
    while the earliest infected node i in infectious_list…
            …has been infectious longer than δ:
        remove i from infectious_list

    # Check if the outbreak died.
    if infectious_list is empty:
        return

    # SI → II code starts here.
    set now_end to the end of infectious_list
    set beginning to the beginning of infectious_list
    while beginning < now_end:
        let i be the node pointed to by beginning
        for all susceptible neighbors j of i:
            with a probability λ:
                add j to the end of infectious_list
                set j's infection time to time
                index_swap(j)
            else:
                increment beginning
```

Fig. 3. Pseudo code for the no-state-label algorithm of the constant duration version. The notations are similar to Fig. 1.

```
# Initialize.
for all vertices i:
      label i susceptible
label source infectious

initialize infectious_list to a list containing only source
initialize source's infection time to 0
for every neighbor i of source:
      add (source,i) to the inactive links list

for time starting from 1:
      # I → R code starts here.
      while the earliest infected node i in infectious_list…
                  …has been infectious longer than δ:
            remove i from infectious_list

      # Check if the outbreak died.
      if infectious_list is empty:
            return

      # SI → II code starts here.
      # The active link list is the one SI links to test
      # are picked from. The passive link list is being
      # filled up for the next time step.
      swap the active and inactive link lists
      for links (i,j) in the active link list:
            # Here we assume the link sorted so that i is infected earlier than j.
            if i is infectious:
                  if j is susceptible:
                        with a probability λ:
                              add j to infectious_list
                              store j's infection time
                              for every neighbor k of j:
                                    if k is susceptible:
                                          append (j,k) to the inactive link list
                  else:
                        append (i,j) to the inactive link list
```

Fig. 4. Pseudo code for the incomplete SI-link list algorithm of the constant duration version. The notations are similar to Fig. 1.

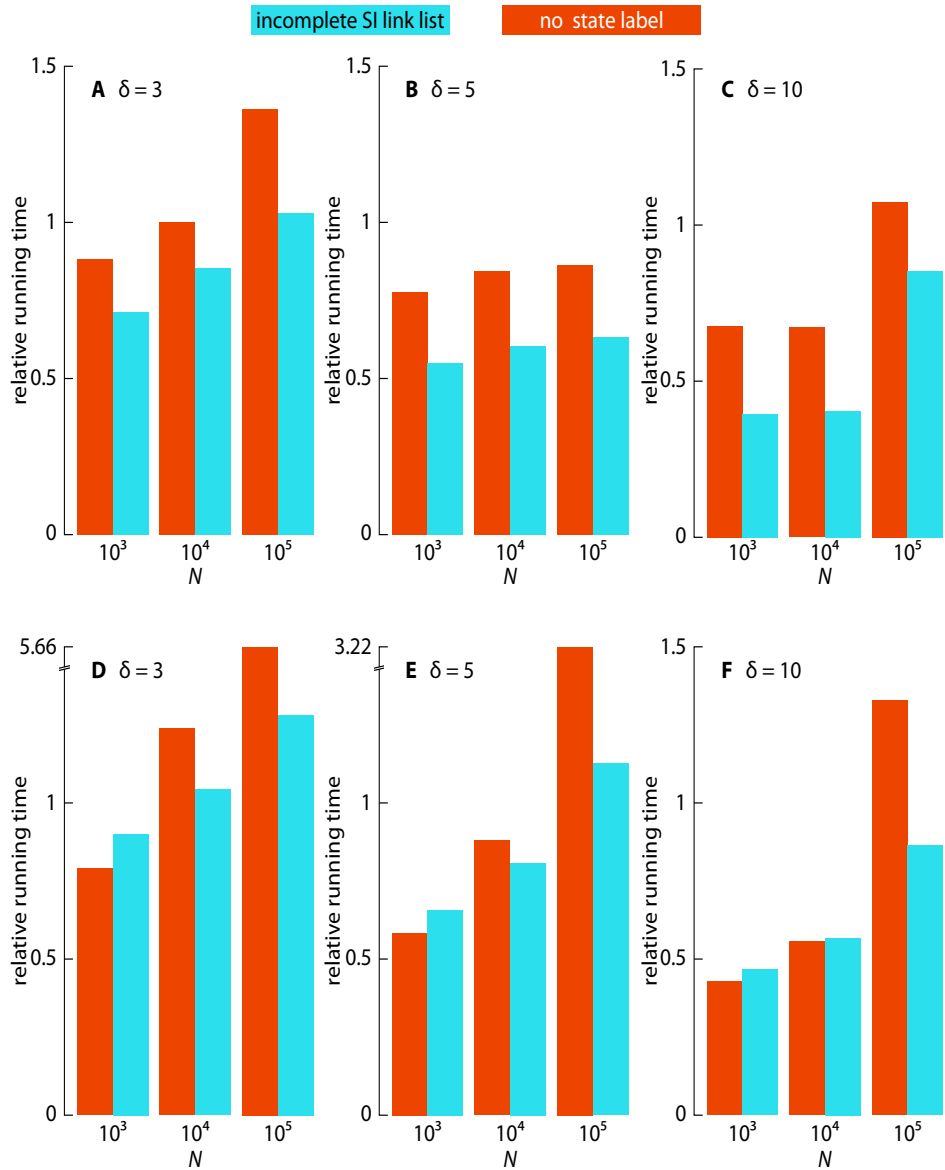

Fig. 5. Running time of the different versions of the constant rate version as a function of system size. A–C show results for random regular graphs with average degree 3. D–F show results for Barabási-Albert networks with average degree 20. The values are averaged over $10^5$ runs.

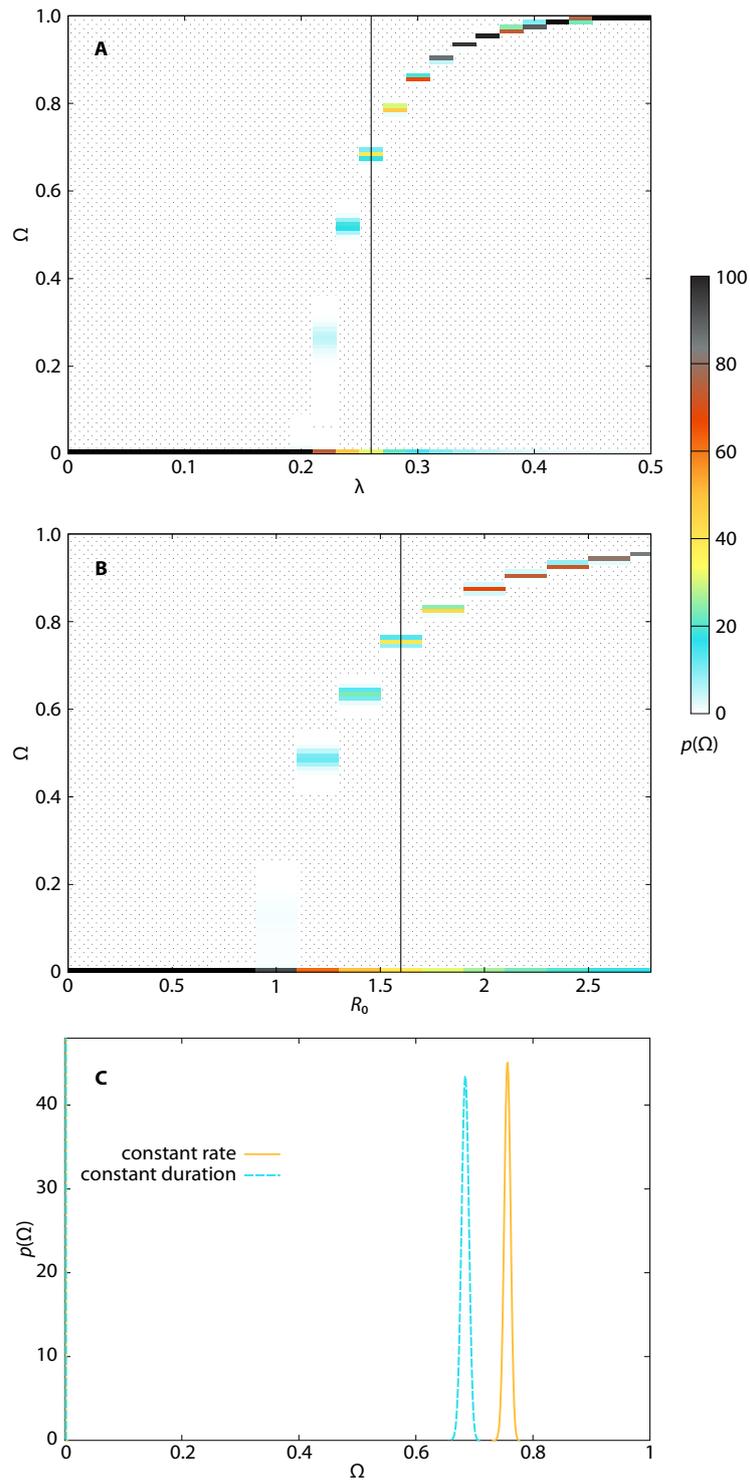

Fig. 6. Outbreak diversity. Panels A and B shows the probability density of the outbreak sizes as functions of the main control parameters ($\lambda$ for the constant duration version, panel A, and $R_0$ for the constant rate version, panel B). Panel C shows a cross section of the plots in panels A and B for the case when the average outbreak size is 0.5. The vertical lines in panels A and B indicate the parameter values in C. The plots are gathered through 1000 network realizations and $10^4$ disease runs per network. The underlying networks are random regular with degree 3.

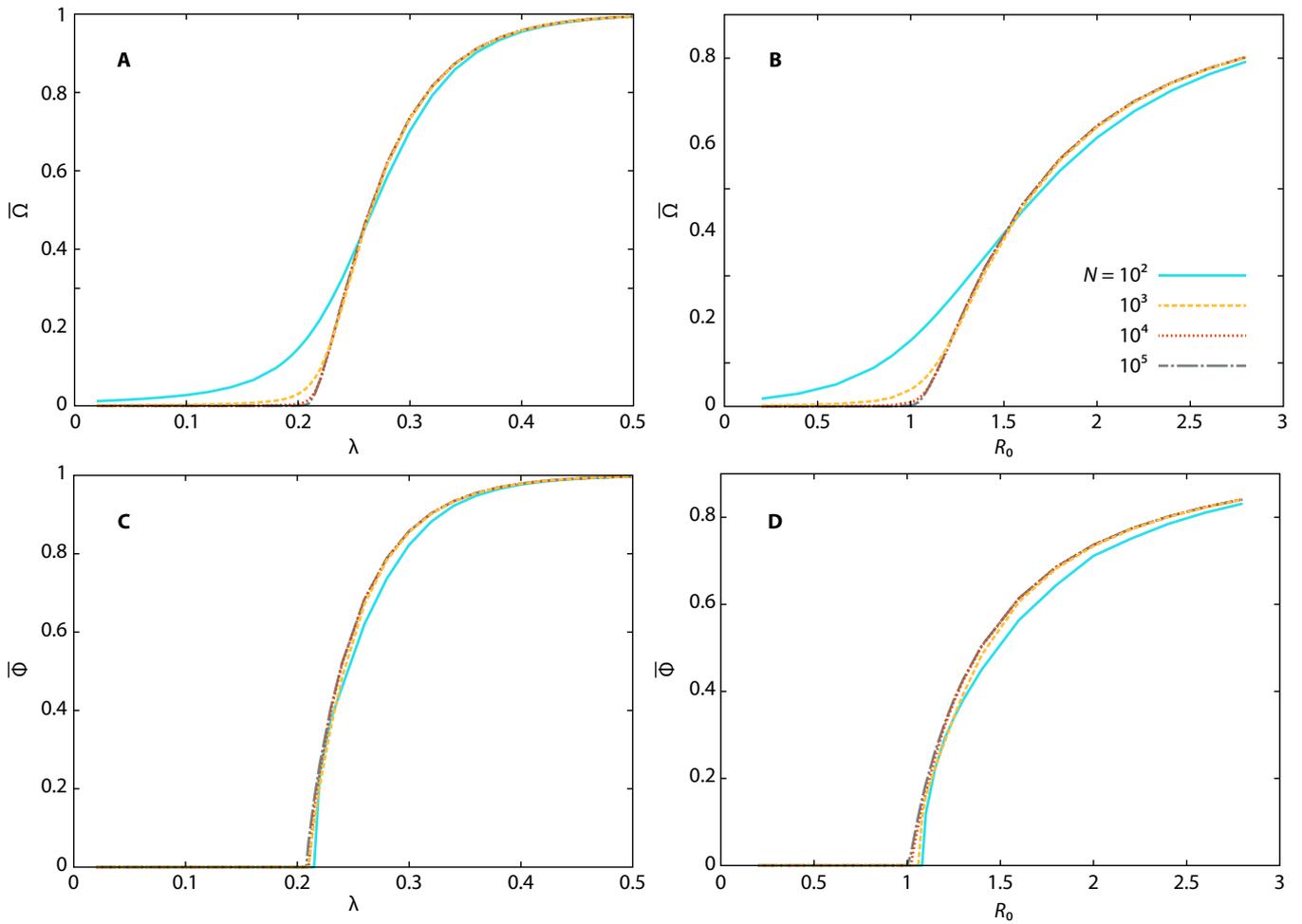

Fig. 7. Average outbreak size (A and B) and outbreak probability (C and D) as a function of the main control parameters ($\lambda$ for the constant duration version and $R_0$ for the constant rate version). Panels A and C show results for the constant duration version and panels B and D are corresponding plots for the constant rate version. The plots are averages over 1000 network realizations and 10000 disease runs per network. The underlying networks are random regular graphs with degree 3.

algorithm until a certain fraction of nodes are infectious and then switch to one of the other ones.

The overall running time that we do not show for simplification scales roughly linearly. We note that the relative running time curves are not so smooth, which must an effect of the hardware and operating system (an iMac assembled 2013 with 32Gb RAM and a four-core Intel i7-3770 processor, running OS X Mavericks).

*Temporal networks*

As mentioned, sometimes one has information about when contacts happen between pairs of individuals, not only what pairs that are connected [2,3]. Typically, such data comes in the form of contact sequences—lists of triples representing id-numbers of the two individuals in contact and the time of the contact. To simulate disease outbreaks on contact sequences is similar to static networks. First, one would keep the contact sequence sorted and run through the contact in time order. Then, simulate an infection event by letting the susceptible become infected with probability $\lambda$ upon an encounter with any infectious other.

One needs to pay some extra attention to contacts happening the same time step. If one just runs the naïve algorithm outlined above, one will effectively assume contacts earlier in the list happen before contacts happening after. One solution is to pick simultaneous contacts in a random order every run of the algorithm. Another version is not to let the disease spread through a contact happening the same time as its infectious node got infected. In either of these versions it is convenient of keeping all the contacts of a time step in a list. Disease spreading is, of course, slightly faster in the first of these versions. Which version to choose is a question about what type of contacts the data set represents and what type of disease one considers.

One slight complication for the constant rate version is that one have to separate $R_0$ into $\beta$ and $\nu$. This will make the program a tad slower (one need more random numbers—on the other hand, random-number generation should not be a bottleneck for any of these algorithms).

*Other versions*

A generalization of the constant duration version would be to sample the disease durations from a (perhaps observed) probability distribution. The code would work much the same, only that one would not be able to just append nodes to the list of infectious nodes, rather one would have to sort them into the list. This sorting could be done by the bisection method (for optimal worst-case time complexity). If the duration distribution is not very broad, it could perhaps be faster to go through the elements sequentially from one side.

We have occasionally seen authors starting the outbreak by letting a fraction of the population, rather than one individual, be the infection source. Note that, for most applications, this is a logical fallacy. In any given population, there must have been a unique, firstly infected individual. If we assume that there is more than one source, then these must be connected through contacts outside of our system. So if the initial configuration were contingent on these other contacts, probably the remainder of the outbreak evolution would also be that. Logically, the hidden network should thus be included which brings us back to a simulation with one infection seed.

## Similarities and differences in the output

*Outbreak diversity*

In Fig. 6A and B, we plot the probability density of the expected outbreak size (fraction of nodes that are Recovered at the end of the simulation) $\Omega$ as functions of the SIR model parameters ($R_0$ for the constant rate version and $\lambda$ for the constant duration version). We see the epidemic threshold manifested very clearly as the probability density branches into a bimodal Ω distribution above the threshold. The outbreak distribution conditioned on the outbreak taking off is very narrow. This means that there is an uncertainty whether or not the disease starts, but other than that the outbreak is very predictable. As we increase the parameters values, outbreaks will be both larger and more frequent. The big picture is very similar for the two versions, which is reassuring for the many studies that use either one without any motivation from the medical literature. The two plots are not identical, however. In Fig. 6C, we show the probability distribution for outbreak sizes with average outbreak size 50%. The constant-rate version has a larger variance with more outbreaks dying early and otherwise larger outbreaks.

*Average outbreak size*

Just like the outbreak diversity, the average Ω values behave qualitatively much the same for the two versions. In Fig. 7A and B, we plot this very common quantity (often called *prevalence*) for characterizing epidemics. In statistical physics, a quantity that indicates two different regions of behavior by being zero in one region and non-zero in the other is called *order parameter*. The point where the change occurs is a *phase transition* (in epidemiological terminology it is called *epidemic threshold*). Another order parameter that shows the phase transition even clearer is the average fraction of outbreaks. It is easily defined by considering a probability distribution like the one in Fig. 6C. If we follow this distribution in increasing parameter values the curve does first decrease (sharply), then increase again. Φ is the fraction of points that are not in the small-outbreak peak (starting counting when $p(\Omega)$ starts to increase again). This quantity is plotted for the two versions in Fig. 7C and D. As the system size increases, the transition looks increasingly clear. For the constant-rate version, the $R_0$ value that marks the transition (the *critical* value, in physics jargon) is one. This is a classic theoretical result [1]. For the constant-duration version, we are not aware of any theory predicting the critical λ value. As seen in Fig. 7, it is around 0.22.

**Discussion**

A main conclusion from this work is that the constant rate and constant duration versions give qualitatively very similar results. The choice between these two versions need not be related to the disease (for example, if one would like to compare the results to analytical calculations, the constant rate version is the standard choice).

The choice of algorithm can make a large difference, but it is very much dependent on the network structure. If speed is important, one could either carefully chose the algorithm for the particular network in question or built the speed testing into a hybrid algorithm. One could probably speed up the running time even more by switching algorithm during the simulation. In the early phase of an outbreak, the canonical algorithm is probably faster (since the chance a neighbor of an infectious is susceptible is larger); in the end of an outbreak no-state-label algorithm could probably be fastest (as it can pick the SI links deterministically).

**Acknowledgments**

This research was supported by Basic Science Research Program through the National Research Foundation of Korea (NRF) funded by the Ministry of Education (2013R1A1A2011947) and the Swedish Research Council.